\documentclass[12pt]{IOP/iopart}
\usepackage{IOP/iopams}
\usepackage{amsmath,amssymb,amsthm}
\usepackage{braket}
\usepackage{cite, hyperref, orcidlink}
\usepackage[nameinlink]{cleveref}
\AddToHook{cmd/appendix/before}{\crefalias{section}{appendix}}
\usepackage{float, graphicx, subcaption}
\usepackage{xcolor}

\begin{document}

\title{Entanglement percolation in random quantum networks}

\author{Alessandro Romancino$^1$\orcidlink{0009-0004-2812-6251}, Jordi  Romero-Pallejà$^2$\orcidlink{0009-0007-2287-8446}, G. Massimo Palma$^1$\orcidlink{0000-0001-7009-4573} and Anna Sanpera$^{2, 3}$\orcidlink{0000-0002-8970-6127}}

\address{$^1$ Dipartimento di Fisica e Chimica ``E. Segrè'', Università degli Studi di Palermo, Via Archirafi 36, 90123 Palermo, Italy}
\address{$^2$ Grup d'Informació Quàntica, Departament de Física, Universitat Autònoma de Barcelona, 08193 Bellaterra, Spain}
\address{$^3$ ICREA, Pg. Lluís Companys 23, 08010 Barcelona, Spain}
\ead{alessandro.romancino@unipa.it}

\begin{abstract}
Entanglement percolation aims at generating maximal entanglement between any two nodes of a quantum network by utilizing strategies based solely on local operations and classical communication between the nodes. As it happens in classical percolation theory, the topology of the network is crucial, but also the entanglement shared between the nodes of the network. In a network of identically partially entangled states, the network topology determines the minimum entanglement needed for percolation. In this work, we generalize the protocol to scenarios where the initial entanglement shared between any two nodes of the network is not the same but has some randomness. In such cases, we find that for classical entanglement percolation, only the average initial entanglement is relevant. In contrast, the quantum entanglement percolation protocol (within the q-swap framework) degrades under these more realistic conditions as the width of the distribution increases, suggesting that Random CEP may become the optimal LOCC strategy in sufficiently heterogeneous quantum networks.
\end{abstract}

\noindent\textit{Keywords}: quantum networks, entanglement, quantum information, quantum internet, percolation, stochastic locc, random states, entanglement distribution

\section{Introduction}
\label{sec:intro}

Despite major advances in quantum technologies, achieving reliable and efficient entanglement distribution across distant nodes of a quantum network (QN) remains one of the most challenging tasks today. Nevertheless, several real-world implementations of quantum networks have recently been demonstrated on various scales \cite{2023hardware, 2022realworld, 2020city, 2020quantnet, 2024thomas}.

Establishing distributed entanglement is a crucial ingredient for the future Quantum Internet \cite{2008kimble, 2018wehner}, as well as for protocols involving conference key agreement (CKA) \cite{2020murta}, quantum federated learning \cite{2025nguyen}, quantum cryptography \cite{1991ekert, 2014bennett}, superdense coding \cite{1992bennett}, secure quantum key distribution \cite{2014lo} and quantum teleportation \cite{1993bennett}. However, due to unavoidable environmental decoherence, the creation of reliable pairwise entanglement at large distances remains a difficult experimental endeavor \cite{satellite, longrange}.

To overcome these limitations, inherent to quantum channels, quantum repeaters have been proposed as a way to achieve perfect pairwise entanglement distribution in one dimensional topologies \cite{1998briegel, 1997cirac,1999dur}. In quantum repeaters, a chain of intermediate nodes act as repeaters of maximally entangled states by applying entanglement swapping \cite{1993zukowski, 1999bose}. Another protocol to achieve maximal entanglement is the so-called entanglement distillation \cite{1996bennet1, 1996bennet2, 1996bennet3, 1996sanpera}, where multiple copies of a partially entangled state are distilled into a smaller number of maximally entangled states. Both methods, however, show fundamental limitations \cite{2017pirandola, 2022winnel}, regarding the action of noise or fidelity. Recently, multiplexing entanglement sharing has also been developed as a way to distribute entanglement \cite{2025ruskuc}.

Here, we will focus on an alternative approach based on percolation on quantum networks called entanglement percolation \cite{2007acin, 2009cuquet, 2008perseguers, 2009lapeyre, 2013perseguers, 2016siomau, 2023review, 2024wang}. A quantum network consists of a set of nodes (users) capable of performing local operations, and the edges (or links) correspond to partially entangled pairs shared between the nodes \cite{2019biamonte}. The goal of entanglement percolation is to establish a maximally entangled state between two arbitrary nodes of a network, when the nodes are connected with partially entangled states. The entanglement transport (percolation) on the network depends non-trivially on the network topology (degree of connectivity of the nodes), the robustness of its architecture \cite{2018das} and the amount of initial entanglement, typically assumed to be homogeneous across all links.

We briefly review the classical and quantum entanglement percolation protocols, which are the crucial concepts regarding percolation in quantum networks. The classical entanglement percolation (CEP) protocol starts from a quantum network whose edges consist of identical partially entangled bipartite states.\footnote{In the original paper \cite{2007acin}, the term “classical” is used purely as an analogy with classical percolation theory; the protocol itself is not classical.}
The strategy is then to “gamble with entanglement” by applying stochastic local operations and classical communication (SLOCC). The effect of SLOCC is that each edge connecting a pair of nodes is converted into a maximally entangled pair with probability $p$; otherwise, the edge is removed from the network with probability $1-p$. This procedure maps the problem onto the classical percolation model familiar from statistical physics \cite{newman}. Every lattice then exhibits a corresponding percolation threshold. 

A more efficient protocol is the so-called quantum entanglement percolation (QEP) one. In this case, each node of the network is preprocessed using some quantum operations, for instance performing entanglement swapping via ``$q$-swaps'' operations \cite{2009cuquet}. In doing so, the original lattice structure is transformed into some other lattice for which the percolation threshold is lower \cite{2007acin}. The change in network topology results in an efficiency bound that can be much better than the original CEP \cite{2009cuquet, 2011wu}. However, the optimal protocol for quantum entanglement percolation on a given network still remains an open question \cite{2013perseguers}, and heuristics can be used to optimize it in some capacity \cite{2025degirolamo}.

To the best of our knowledge, entanglement percolation has been explored using pure bipartite qubits \cite{2007acin}, mixed states \cite{2009mixed, 2026wang}, multipartite entangled states \cite{2010perseguers, 2024khanna}, noisy channels \cite{2024oh}, continuous-variable Gaussian states \cite{2025zhao}.

While in all the above cases the initial entanglement is typically assumed to be homogeneous across all links, realistic quantum communication frameworks exhibit inherent heterogeneity. Physical edges possess varying lengths, leading to distance-dependent exponential photon loss in optical fibers, which yields unequal entanglement fidelities across the network \cite{2018wehner}. Additionally, the integration of distributed quantum memories \cite{2025meng} subjects the stored quantum states to variable, link-dependent coherence times and fluctuating operational efficiencies. Furthermore, the presence of disorder in interaction-driven network Hamiltonians inherently generates heterogeneous entanglement distributions across the lattice \cite{2026cirigliano}.

These physical constraints dictate that the shared entanglement across all edges follows a statistical distribution rather than a uniform constant. Here, we focus on the efficiency of entanglement percolation for the more realistic scenario in which the degree of entanglement corresponding to the edges is randomly distributed.

The structure of the article is as follows: in \Cref{sec:ent-perc}, we introduce the main concepts underlying the entanglement percolation protocol as originally proposed in Acín \etal.~\cite{2007acin}. We then review the basis of the  CEP protocol and the QEP protocol. In \Cref{sec:rand}, we extend the previous analysis to networks with nonidentical shared initial states. We present our results for CEP in \Cref{sec:rand-cep} and for QEP in \Cref{sec:rand-qep}. In \Cref{sec:conc}, we summarize our findings and open questions. For completeness, a brief overview of classical percolation theory is presented in \Cref{sec:comp_net} while some formulas and results from LOCC theory are outlined in \Cref{sec:quant_info}.

\section{Entanglement percolation in quantum networks}
\label{sec:ent-perc}

\begin{figure}
    \centering
    \includegraphics[width=0.4\textwidth]{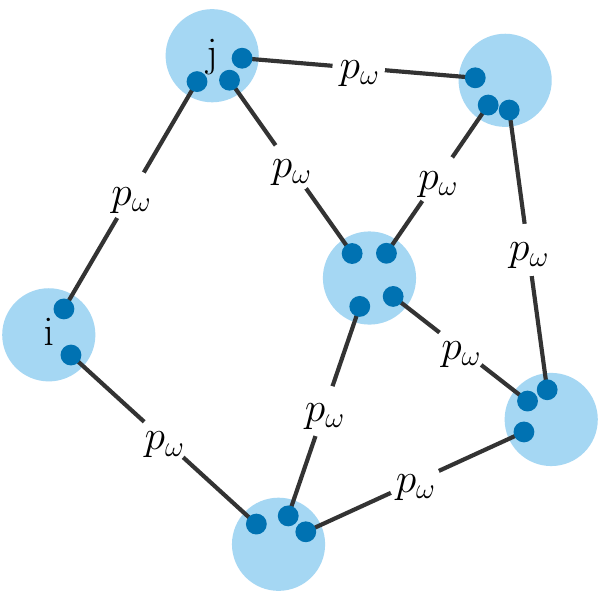}
    \caption{Example of a quantum network consisting of identical copies of state $\ket{\omega}$ with Singlet Conversion Probability (SCP) $p_{\omega} = 2\lambda^\omega_2$. Nodes $i$ and $j$ are highlighted.}
    \label{fig:qn}
\end{figure}

A quantum network can be modeled as a graph $G=(V,E)$, where the vertices $V$ represent nodes (users) capable of local quantum operations, and the edges $E$ represent bipartite quantum states shared between pairs of nodes. Each edge is associated with a partially entangled two‑qubit state. In \Cref{fig:qn}, a network whose edges consist of partially entangled two-qubit states, $\ket{\omega}$, is shown. Users are allowed to use local operations and classical communication (LOCC) to achieve singlets $\ket{\Psi^-}$, between two arbitrary nodes of the quantum network. Using the Schmidt decomposition of a shared bipartite pure state $\ket{\omega} = \sqrt{\lambda_1^{\omega}} \ket{00} + \sqrt{\lambda_2^{\omega}} \ket{11}$ (with $\lambda_2^{\omega} \leq \lambda_1^{\omega}$), together with stochastic majorization \cite{1999vidal}, it can be shown that any such state can be transformed into a singlet (or equivalently into any maximally entangled Bell state) by means of LOCC with the singlet conversion probability (SCP) $p_{\omega}= \min \{ 1, 2\lambda_2^\omega \}$. See \Cref{sec:quant_info} for the derivation details.

\begin{figure}
    \centering
    \includegraphics[width=0.5\textwidth]{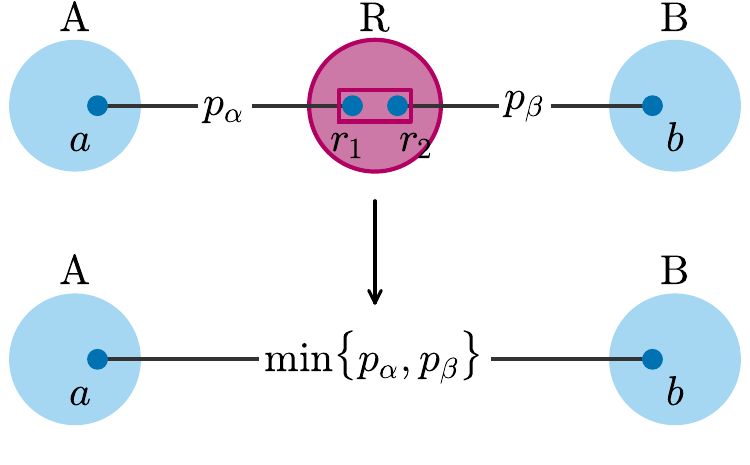}
    \caption{Representation of an entanglement swapping procedure, where a joint Bell measurement is performed on systems $R_1$ and $R_2$.}
    \label{fig:ent-swap}
\end{figure}

To achieve entanglement distribution, one makes use of only two types of local operations and classical communication protocols, namely SLOCC purification and entanglement swapping.
\begin{itemize}
    \item \textbf{SLOCC purification:} nodes $i$ and $j$ are connected by a partially entangled state $\ket{\omega}_{ij}$ (like in \Cref{fig:qn}). The optimal SLOCC purification protocol is applied on the qubits of nodes $i$ and $j$. This operation, as explained in \Cref{sec:quant_info}, converts the state $\ket{\omega}_{ij}$ into the singlet $\ket{\Psi^-}_{ij}$ with finite probability $p_\omega$ or, if unsuccessful, into a product state with probability $1 - p_\omega$.
    \item \textbf{Entanglement swapping:} as shown in \Cref{fig:ent-swap}, two nodes $A$ and $B$ are connected by an intermediate node R, sharing states $\ket{\alpha}_{ar_1}$ and $\ket{\beta}_{br_2}$. Then, a joint Bell measurement is performed on node $R$ by measuring systems $r_1$ and $r_2$. This creates a new entangled state between $A$ and $B$, while the node $R$ is removed. Considering that the singlet conversion probabilities (SCPs) are $p_\alpha$ and $p_\beta$ respectively, one ends up with a state with an SCP of $p_\text{swap} = \min\{p_\alpha, p_\beta\}$, also called ``entanglement of single pair purification'' \cite{1999bose, 2023entswap}. In this section we consider just the case where $p_\alpha = p_\beta = p$ for all the links. However, as we will see, the role of different $p_i$ will, instead, be crucial in \Cref{sec:rand-qep}.
\end{itemize}

\subsection{Classical Entanglement Percolation}
\label{sec:cep}

\begin{figure}
    \centering
    \begin{subfigure}[b]{0.35\textwidth}
        \includegraphics[width=\textwidth]{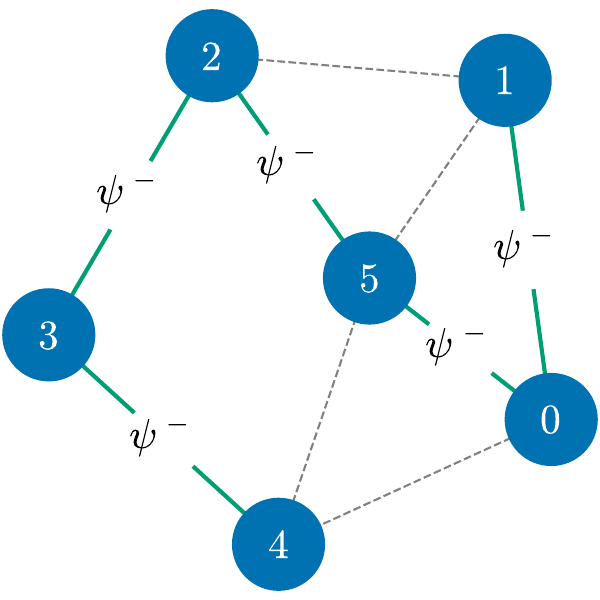}
        \caption{System after the SLOCC purifications.}
        \label{fig:qnrun-1}
    \end{subfigure}
    \hspace{3em}
    \begin{subfigure}[b]{0.35\textwidth}
        \includegraphics[width=\textwidth]{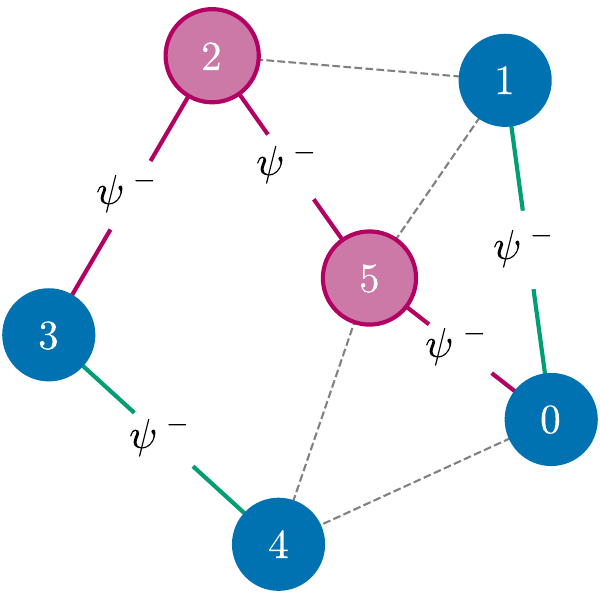}
        \caption{The path of singlets is highlighted.}
        \label{fig:qnrun-2}
    \end{subfigure}
    \caption{A single (successful) run of a classical entanglement percolation protocol. The final singlet is created between node 0 and node 3 after applying entanglement swapping on nodes 2 and 5.}
    \label{fig:qn-run}
\end{figure}

The Classical Entanglement Percolation (CEP) protocol \cite{2007acin} consists of the following procedure:
\begin{enumerate}
    \item The optimal SLOCC purification operation is applied to each link in the network. The links are either successfully converted into a singlet with probability $p_\omega$ or removed from the network (\Cref{fig:qnrun-1}). To perform this LOCC operation the nodes must have local knowledge of the exact Schmidt coefficients of the state they share (see \Cref{sec:quant_info}).
    \item If the initial SCP is high enough, a chain of singlets connecting the desired nodes is created. Identifying this path requires classical communication across the whole network to determine which links were successfully converted. Once the path is established, entanglement swapping is applied in the intermediate nodes, generating the final desired singlet, as shown in \Cref{fig:qnrun-2}.
\end{enumerate}
The classical theory of percolation states that if the initial $p_\omega$ is high enough, a giant connected component emerges. Consequently there exists a strictly positive probability of finding a path connecting two arbitrarily chosen nodes. This means that for entanglement percolation to be successful, it is required for the initial states to have an initial SCP $p_\omega > p_c$, where $p_c$ is the percolation threshold of the original network \cite{2007acin, 2009cuquet}.

This threshold is called Classical Entanglement Percolation (CEP), in analogy with the classical theory of percolation \cite{2013perseguers, 2009lapeyre, 2008perseguers, 2016siomau}. For example, a square lattice quantum network will require states with $p_\omega > p^\text{square}_c = 1/2$ \cite{1980kesten}, that is, $p^\text{CEP}_\text{square} = 1/2$. The percolation threshold depends only on the network topology, and different types of complex networks can show very significant differences in the percolation threshold as shown in \cite{newman, latora, 2009cuquet, 2011wu}.

\subsection{Quantum Entanglement Percolation}
\label{sec:qep}

\begin{figure}
    \centering
    \begin{subfigure}[b]{0.35\textwidth}
        \includegraphics[width=\textwidth]{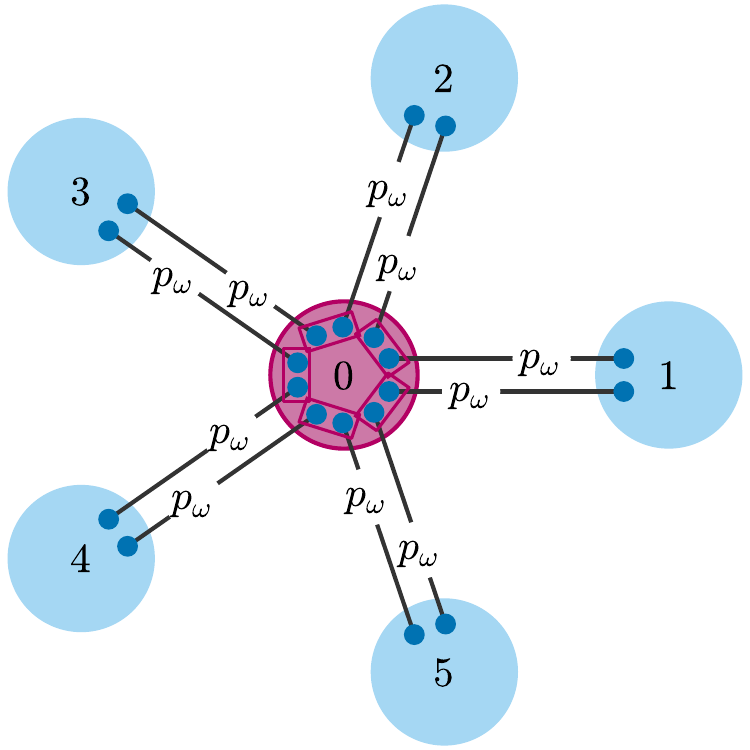}
        \caption{Initial start graph $S_5$.}
    \end{subfigure}
    \hspace{3em}
    \begin{subfigure}[b]{0.35\textwidth}
        \includegraphics[width=\textwidth]{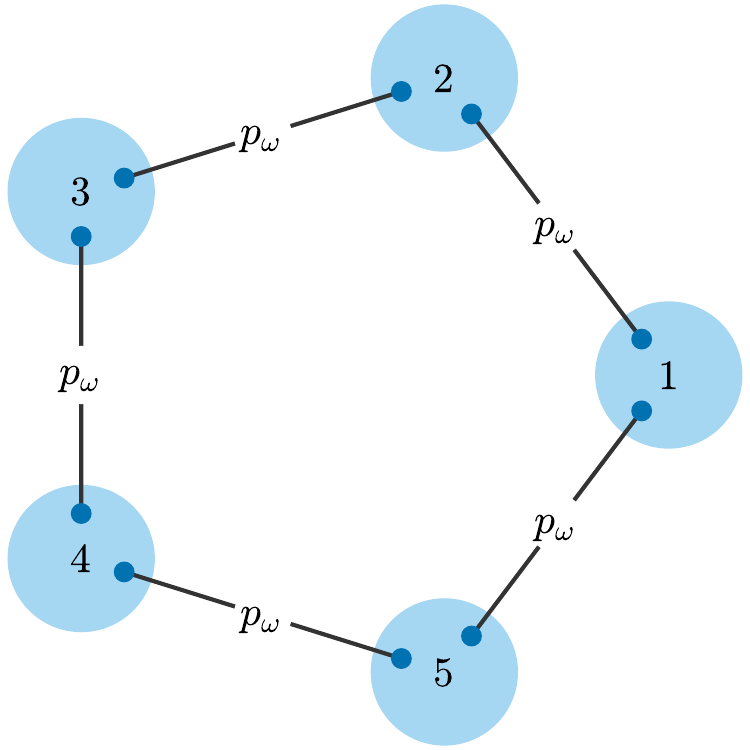}
        \caption{After applying the $5$-swap.}
    \end{subfigure}
    \caption{An example of a $5$-swap operation. (a) Initial star graph $S_5$ with a central node $0$ connected with two identical states $\ket{\omega}$ with each of the other nodes. (b) End result after performing the $5$-swap between the $0$ node and all other nodes. This creates new entangled states between the external nodes and erases node $0$ from the network.}
    \label{fig:qswap}
\end{figure}

It has been shown that the CEP protocol is not optimal for many Quantum Networks \cite{2007acin, 2009cuquet}. In those cases, a better performance can be achieved with the Quantum Entanglement Percolation (QEP) protocol. The starting setup for this protocol consists of a ``multigraph quantum network'', that is, a quantum network where each pair of nodes can share multiple bipartite states (see \Cref{sec:multi} for a detailed explanation) and implement a procedure called ``$q$-swap''. It consists of applying entanglement swapping operations in nodes at the center of a star subgraph $S_q$ transforming it into a $C_q$ cycle, as can be seen in \Cref{fig:qswap}. These operations change the topology of the network, modifying the percolation threshold drastically \cite{2009cuquet, 2011cuquet}.

Therefore, in the QEP protocol, first the network is preprocessed by applying appropriate $q$-swaps, and then the CEP protocol is applied on the new network.

\begin{figure}
    \centering
    \begin{subfigure}[b]{0.35\textwidth}
        \includegraphics[width=\linewidth]{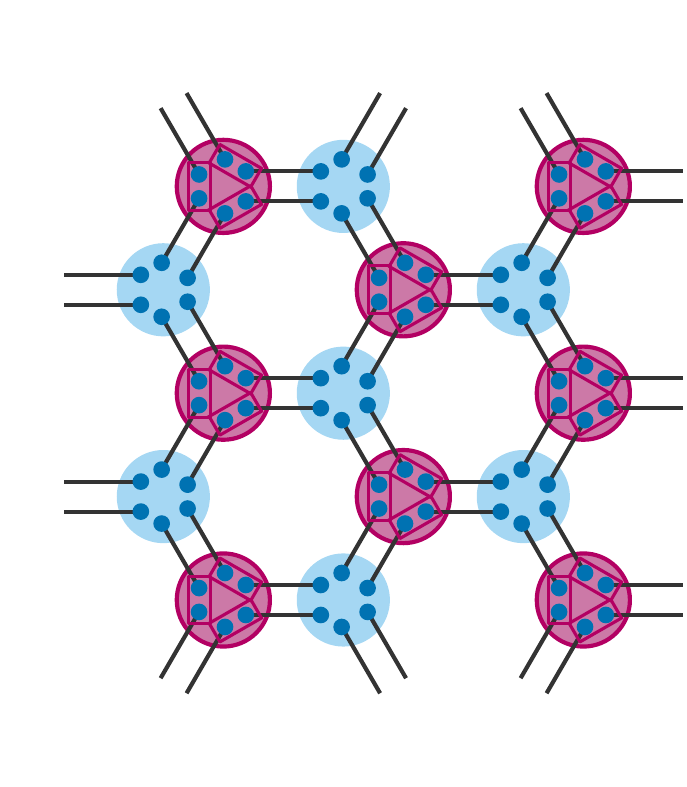}
        \caption{Starting network.}
    \end{subfigure}
    \hspace{3em}
    \begin{subfigure}[b]{0.35\textwidth}
        \includegraphics[width=\linewidth]{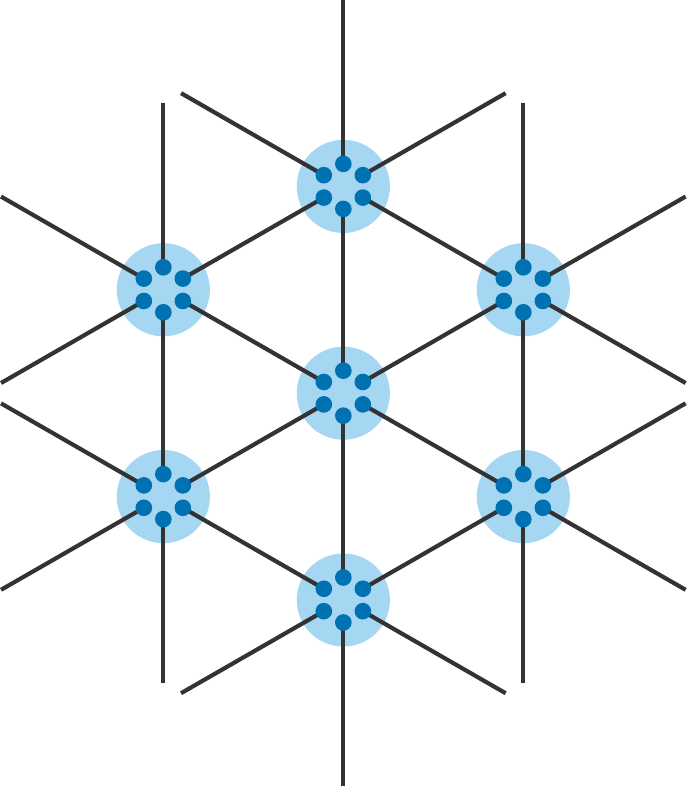}
        \caption{After the $3$-swaps.}
    \end{subfigure}
    \caption{Representation of the initial and final networks after applying the QEP protocol to a double-bond honeycomb lattice.}
    \label{fig:hextri}
\end{figure}

As a paradigmatic example \cite{2007acin}, starting with a double-bond honeycomb lattice, as shown in  \Cref{fig:hextri}, by applying the CEP protocol, knowing that $p^\text{hexagon}_c = 1 - 2 \sin (\pi/18)$ \cite{1964sykes}, one obtains
\begin{equation}
    \label{eq:hex}
    p^\text{CEP} = 2 \left[1 - \sqrt{\frac{1}{2} + \sin\left(\frac{\pi}{18}\right)}\right] \approx 0.358
\end{equation}
If, instead, we apply the QEP protocol, one initially performs $3$-swaps and ends up with the triangular lattice, as shown in \Cref{fig:hextri}, which has $p^\text{triangle}_c = 2 \sin (\pi/18)$ \cite{1964sykes}. This means that
\begin{equation}
    \label{eq:tri}
    p^\text{QEP} = p^\text{CEP}_\text{triangle} = 2 \sin (\pi/18) \approx 0.347
\end{equation}
In the case here illustrated, the difference between CEP and QEP is small, but the QEP protocol can yield much better improvements in complex quantum networks \cite{2009cuquet, 2011cuquet}.

It is also known that QEP is not optimal. Given an arbitrary network, the order of the entanglement swapping and stochastic purification process can be tailored to the specific network, going beyond the QEP performance. It is still unknown if, given any quantum network, there exists a lower bound for initial entanglement below which it is impossible to entangle two distant qubits using only LOCC \cite{2013perseguers}.

\section{Entanglement percolation in random quantum networks}
\label{sec:rand}

Entanglement percolation protocols have been extended to networks of multipartite states \cite{2010perseguers} and to networks of mixed states \cite{2009mixed}. Here, we analyze a still unexplored but highly relevant and realistic scenario: a random quantum network (RQN). In this setting, we relax the usual assumption that all initially shared quantum states are identical and instead assume a statistical distribution of the initial SCPs. Such statistical distributions naturally emerge in physical systems, for example, through spatially correlated disorder driven by unitary evolution under disordered network Hamiltonians \cite{2026cirigliano}.

\begin{figure}
    \centering
    \begin{subfigure}[b]{0.35\textwidth}
        \includegraphics[width=\textwidth]{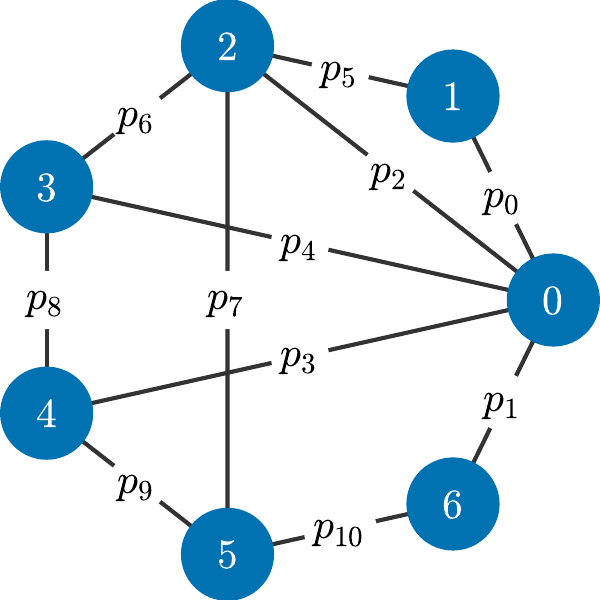}
        \caption{Example of a random state network.}
        \label{fig:random-network}
    \end{subfigure}
    \hspace{3em}
    \begin{subfigure}[b]{0.4\textwidth}
        \includegraphics[width=\textwidth]{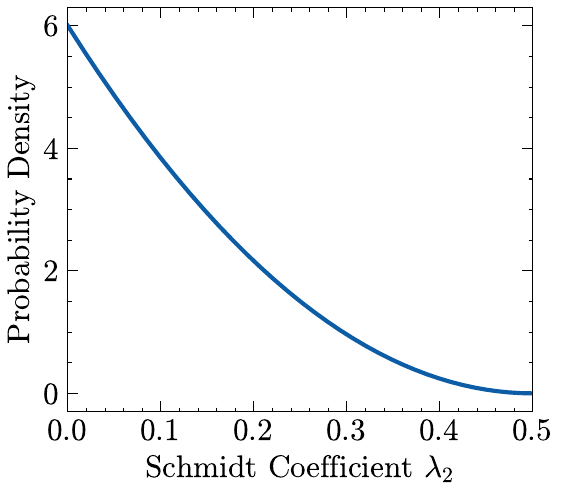}
        \caption{Distribution of the $\lambda_2$ of random Haar states of two qubits.}
        \label{fig:schmidt}
    \end{subfigure}
    \caption{Examples for a random state network. The Haar random states make up a particular distribution from all the possible ones.}
\end{figure}

Let the $k$-th edge state be $\ket{\psi_k}$, the SCP of the $k$-th edge will be $p_k = 2\lambda_2^{\psi_k}$ where $\lambda_2^{\psi_k}$ is the smallest Schmidt coefficient of the state $\ket{\psi_k}$. In the following, we assume that $p_k$ are randomly distributed, as illustrated in \Cref{fig:random-network}. We stress here that such a situation represents a more generic case than just considering Haar-random pure states of two qubits \cite{whatis, 2011zyc}. Notice that Haar states are associated with a unique distribution of Schmidt coefficients \cite{1988lloyd, 1998hall}, meaning that Haar distributed states are just one particular example of all possible distributions of Schmidt coefficients.

In particular, the distribution shown in \Cref{fig:schmidt} of the smallest Schmidt coefficients, $\lambda_2$, of a Haar random state of a pair of qubits, is
\begin{equation}
    p(\lambda_2) = 6 (1 - 2\lambda_2)^2
\end{equation}
which leads to $\langle \lambda_2 \rangle = \frac{1}{8}$. This means that the average SCP of a random Haar state is $\langle p_\text{Haar} \rangle = \frac{1}{4}$.

\subsection{CEP in RQN}
\label{sec:rand-cep}

The implementation of the CEP protocol in these new random quantum networks is straightforward: the optimal SLOCC is applied to each edge (each of which will now have a different probability of success) and then, if the initial amount of entanglement is high enough, a giant cluster of singlets emerges, establishing connectivity at distance.

It is assumed that the realized link SCPs $p_k$ are known locally by the two nodes sharing the respective edge, as this information is strictly necessary for the optimal SLOCC protocol (as shown for the Procrustean method in \Cref{sec:quant_info}).

As an example, a uniform distribution for the SCPs has been assumed
\begin{equation}
    \chi(p) =
    \begin{cases}
        \frac{1}{b - a} &\quad \text{for}\ a \leq p \leq b \\
        0 &\quad \text{elsewhere}
    \end{cases}
    \label{eq:uniform}
\end{equation}
with average value of $\langle p \rangle = \frac{a + b}{2}$ and width $w = b - a$ with $0 \leq a < b \leq 1$. Several random quantum network CEPs have been simulated, with the width $w$ and the average value $\langle p \rangle$ serving as control parameters within their respective admissible bounds. For each parameter set, multiple realizations of the same network -- specifically, a $100 \times 100$ square lattice -- were generated and simulated. Because the edge probabilities are assigned as independent and identically distributed random variables, each realization of the network corresponds to a different random assignment of SCP values to the edges. Accordingly, the Percolation Strength $P_{\infty}$ (i.e., the probability that a randomly chosen node belongs to the largest connected cluster; see \Cref{sec:comp_net}) was estimated by averaging over many independent realizations of the network.

\begin{figure}
    \centering
    \includegraphics[width=0.5\textwidth]{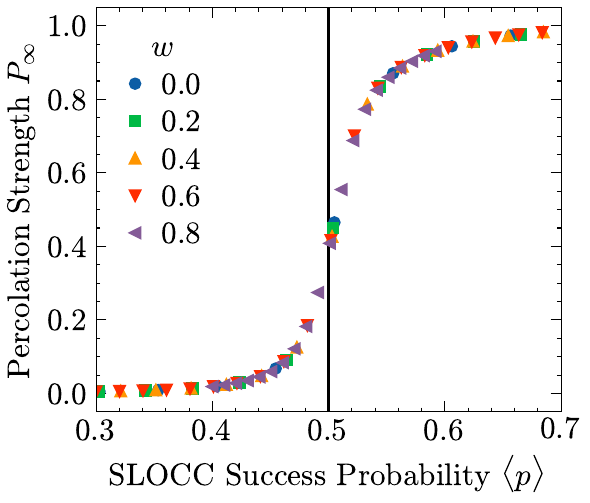}
    \caption{Simulation for the entanglement percolation in a random quantum network. The percolation strength is shown as the order parameter for the transition by changing the SLOCC success probability $\langle p \rangle = \frac{a + b}{2}$. The simulation was done in a $100 \times 100$ square lattice, the percolation threshold $p^\text{square}_c = 0.5$ is shown as a black vertical line. The simulation was done for $w = 0, 0.2, 0.4, 0.6, 0.8$. Note that the plotted domain for each curve is restricted to its mathematically admissible parameter range $w/2 \leq \langle p \rangle \leq 1 - w/2$; for instance, the data points for $w=0.8$ are strictly confined to $\langle p \rangle \in [0.4, 0.6]$.}
    \label{fig:simulation}
\end{figure}

As shown in \Cref{fig:simulation}, the results obtained for the different simulations are identical width-wise. This implies that only the average value of the distribution is relevant for CEP, or, in other words, that only the average initial entanglement is important. We will call this new CEP threshold, Random CEP (RCEP)
\begin{equation}
    p^\text{RCEP}(\{p_i \}) = p^\text{CEP}(\langle p \rangle ) = p_c
    \label{eq:random-cep}
\end{equation}
Therefore, for a sufficiently large system, if the average initial SCP is above the classical percolation threshold $p_c$ then the system percolates, otherwise it does not. An analytical demonstration of this is present in \Cref{sec:rcep-theory}.

\subsection{QEP in RQN}
\label{sec:rand-qep}

\begin{figure}
    \centering
    \begin{subfigure}[b]{0.35\textwidth}
        \includegraphics[width=\textwidth]{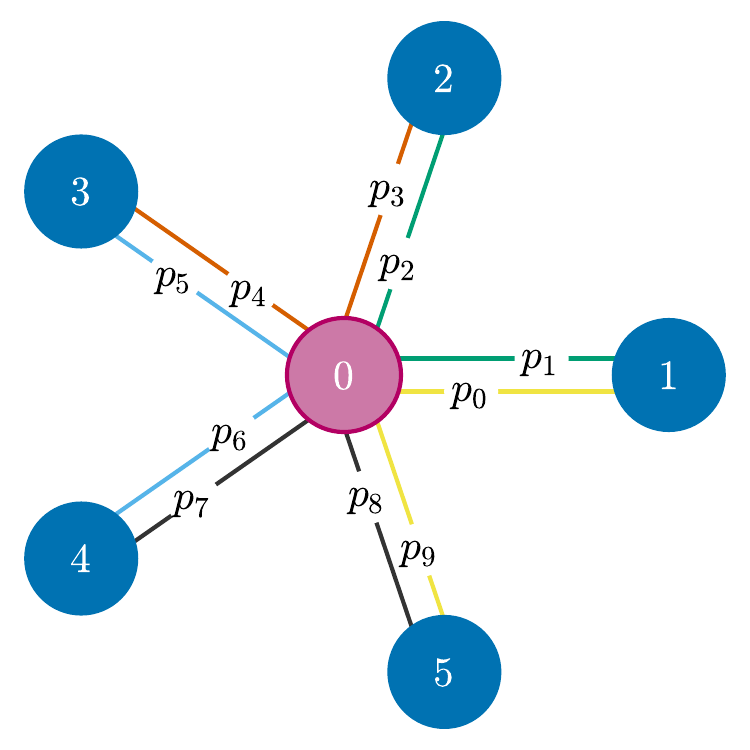}
        \caption{Before applying the $5$-swap.}
    \end{subfigure}
    \hspace{3em}
    \begin{subfigure}[b]{0.35\textwidth}
        \includegraphics[width=\textwidth]{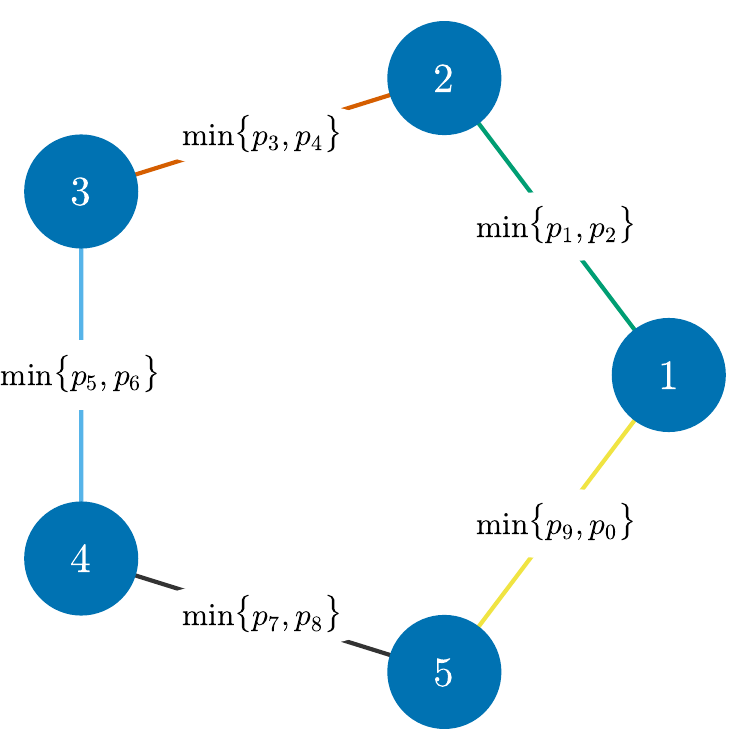}
        \caption{After applying the $5$-swap.}
    \end{subfigure}
    \caption{A $5$-swap applied to a random state network with random multiedges. Because each swap utilizes distinct edges from the initial multigraph, the new edges created after the $q$-swap remain statistically independent random variables.}
    \label{fig:qswap-rand}
\end{figure}

In this section, we compare the efficiency of the QEP protocol and the CEP protocol in random quantum networks. Nevertheless, the $q$-swap defined in \Cref{sec:qep} has to be generalized before proceeding with the protocols comparison. A known result is that entanglement swapping between two states with SCPs $p_{k_1}$ and $p_{k_2}$ yields a new state with an SCP equal to $\min\{p_{k_1}, p_{k_2}\}$ \cite{1999bose}. This means that, in the RQN scenario, the $q$-swaps not only modify the topology of the network, but also degrade the distribution of the SCPs \cite{2026xing}, as shown in \Cref{fig:qswap-rand}.

In the new random quantum network after the $q$-swaps operations the distribution of the SCPs corresponds to the distribution of the minimum of two independent and identically distributed variables (IID). It is important to note that since the initial setup assumes a multigraph (such as the double-bond honeycomb), each entanglement swapping operation within the $q$-swap utilizes distinct, independent edges from the initial multiedges. Consequently, the newly created edges in the transformed lattice remain strictly independent random variables.

This concept of the ``distribution of the minimum'' is common in the field of extreme value theory \cite{evt}. In particular, given $Y = \min\{X_1, X_2\}$ and using the cumulative distribution function (CDF) $F_\text{min}$
\begin{equation}
    F_\text{min}(x) = P(Y \leq x) = P(\min\{X_1, X_2\} \leq x) = 1 - P(X_1 > x, X_2 > x)
\end{equation}
due to the fact that only one $X_i \leq x$ is required. However, the $X_i$ are IIDs and, therefore, $P(X_1 > x, X_2 > x) = [1 - F(x)]^2$ where $F(x) = P(X_i \leq x)$. This leads to a formula for the probability density function (PDF)
\begin{equation}
    f_\text{min}(x) =  2f(x) [1 - F(x)]
    \label{eq:distr-min}
\end{equation}
For example, consider two IID variables drawn from the uniform distribution in \Cref{eq:uniform}. Using \Cref{eq:distr-min}, we obtain the following distribution
\begin{equation}
    \chi_{min}(p) =
    \begin{cases}
        \dfrac{2}{b - a} \left(1 - \dfrac{p - a}{b - a}\right) &\quad \text{for}\ a \leq p \leq b \\
        0 &\quad \text{elsewhere}
    \end{cases}
    \label{eq:uniform-min}
\end{equation}
with average value
\begin{equation}
    \langle p_\text{min} \rangle = \langle p \rangle - w / 6
    \label{eq:minimum}
\end{equation}
This is a reasonable result. The wider the uniform distribution is, the more skewed to the left the distribution of the minimum will be, yielding a lower average.

\begin{figure}
    \centering
    \begin{subfigure}[b]{0.35\textwidth}
        \includegraphics[width=\textwidth]{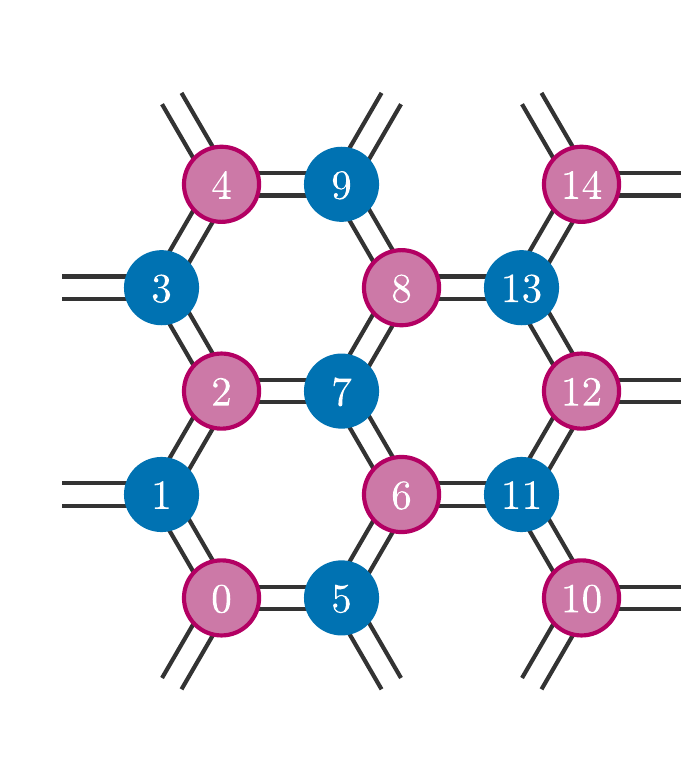}
        \caption{Honeycomb lattice before applying the $q$-swaps.}
    \end{subfigure}
    \hspace{3em}
    \begin{subfigure}[b]{0.35\textwidth}
        \includegraphics[width=\textwidth]{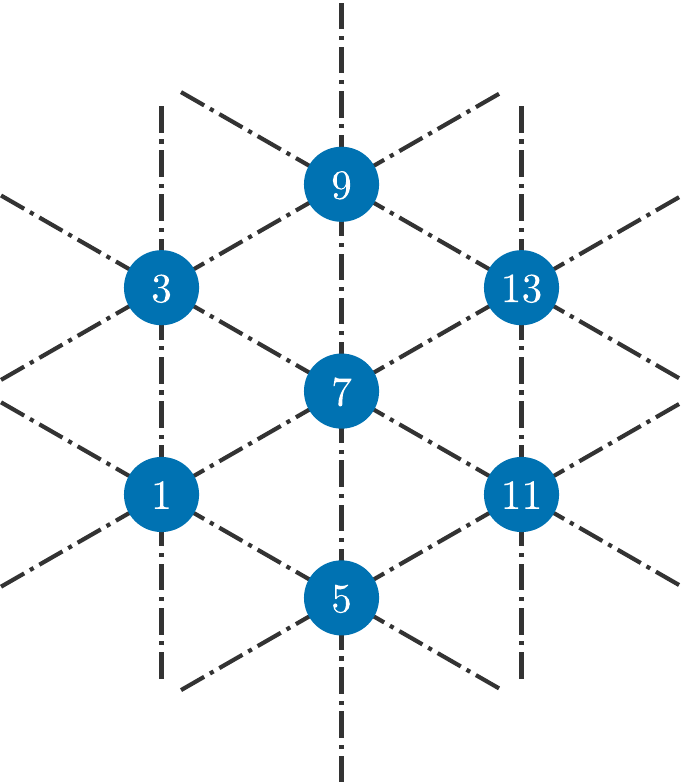}
        \caption{Triangular lattice after applying the $q$-swaps.}
    \end{subfigure}
    \caption{Honeycomb to triangular lattice transformation obtained after applying the $q$-swaps in the nodes highlighted in red. The edges in the newly formed triangular lattice have the SCP following the distribution of the minimum (shown as dotted line in the figure).}
    \label{fig:random-hextri}
\end{figure}

We now analyze the example of a double-bond honeycomb network (like the one in \Cref{sec:qep}) with uniformly distributed SCPs. 
Combining \Cref{eq:random-cep} and $\langle p_{k_1 \otimes k_2} \rangle = p^\text{hexagon}_c$ (see  \Cref{sec:rand-multi} for more details), the RCEP for this network becomes 
\begin{equation}
    p^\text{RCEP} = p^\text{CEP}(\langle p \rangle ) \approx 0.358
\end{equation}
Applying instead the Random QEP (RQEP) protocol, i.e.,  transforming the double-bond honeycomb lattice into a triangular lattice with the distribution of the minimum, we obtain that $\langle p_\text{min} \rangle = p^\text{triangle}_c$. Using \Cref{eq:minimum}, it can be seen that
\begin{equation}
    p^\text{RQEP} = p^\text{QEP}(\langle p \rangle ) + \frac{w}{6} \approx 0.347 + \frac{w}{6}
    \label{eq:randhextri}
\end{equation}
which now depends on the width of the distribution.

\Cref{eq:randhextri} shows that increasing the width worsens the performance of the RQEP protocol, whereas the RCEP threshold does not depend on the width. This means that, for this lattice and distribution, there exists a threshold $w^*$ after which $p^\text{RCEP} < p^\text{RQEP}$. For instance, for $w < w^* \approx 0.067$ the RQEP protocol is more efficient than RCEP for the studied lattice while for $w > w^*$ the protocol is instead disadvantageous.

We note that this behavior is actually quite general. For any continuous distribution with mean $\mu$ and CDF $F(p)$, the expected value of the minimum of two independent draws $Y = \min\{X_1, X_2\}$ is given by
\begin{equation}
    \langle Y \rangle = \mu - \int_{-\infty}^{\infty} F(p) [1 - F(p)] dp
\end{equation}
We now consider distributions belonging to a location-scale family. For these distributions, the shape is fixed and we just shift the mean or stretch the width. The CDF can be written as $F(p) = G\left(\frac{p - \mu}{\sigma}\right)$, where $G(z)$ is the standard baseline shape, $\mu$ is the mean, and $\sigma$ is proportional to the standard deviation.

If we plug this $F(p)$ into our integral and do a simple change of variables $z = \frac{p - \mu}{\sigma}$ (which means $dp = \sigma dz$), we get
\begin{equation}
    \langle p_\text{min} \rangle = \mu - \sigma \int_{-\infty}^{\infty} G(z) [1 - G(z)] dz
\end{equation}
Since $G(z)$ is a valid CDF, the integral over all space just evaluates to a strictly positive constant $C$ that depends only on the standard shape of the distribution, completely independent of $\mu$ and $\sigma$. Mapping $\mu$ back to our average SCP $\langle p \rangle$, we end up with
\begin{equation}
    \langle p_\text{min} \rangle = \langle p \rangle - C\sigma
\end{equation}

This means that if the network's randomness can be modeled by stretching a standard distribution shape, the drop in the expected minimum scales perfectly linearly with the standard deviation.

Therefore, for this general class of distributions, while RCEP is invariant under changes in the width of the distribution, the expected minimum entering the RQEP protocol strictly degrades as a linear function of $\sigma$. Hence, whenever the (non-random) QEP advantage is finite and the admissible range of widths is sufficiently large, a crossover variance exists beyond which RCEP outperforms RQEP.

For this reason, we conjecture that RCEP is the optimal LOCC protocol for entanglement distribution when the variance of the input SCP distribution is sufficiently large.

\section{Conclusions}
\label{sec:conc}

We have generalized the entanglement percolation protocol in quantum networks to the more realistic scenario in which the degree of entanglement between nodes is randomly distributed. Assuming a distribution of states with random Schmidt coefficients, we have found that only the average value of the initial entanglement is important for random classical entanglement percolation (RCEP), mapping again the problem to the classical percolation transition. By applying, instead, the quantum entanglement percolation protocol we found that random quantum networks yield different results than before. We have shown that, in this realistic scenario (RQEP), the efficiency of quantum local operations degrades as the width of the input SCP distribution increases, while RCEP depends only on the mean SCP. As a consequence, for sufficiently broad distributions, RCEP can outperform the corresponding RQEP protocol. We conjecture that, in sufficiently high-variance random quantum networks, RCEP is the optimal LOCC protocol for entanglement distribution.

Applying this framework to the multipartite case might be promising, as the non-random model shows a big improvement over the bipartite case. This is done by mapping the hypergraph representation into a bipartite graph and then analyzing node percolation of the said graph \cite{2010perseguers}. A true full hypergraph percolation theory is still being investigated \cite{2023bianconi}. Another approach, called concurrence (and negativity \cite{2026zhao}) percolation \cite{2021meng, 2022malik, 2023meng, 2025nath} has also been showing promising results and this generalization can be applied to that framework as well. Still, whether there exists a minimum amount of initial entanglement needed in order to achieve a perfect long-distance entanglement is still unknown.

\section*{Data availability}
The code and data that support the findings of this study are openly available on GitHub at \url{https://github.com/alex180500/ent-perc-rand}.

\ack
The authors thank Lorenzo Cirigliano for bringing related work to our attention. The authors thank Enrico Di Benedetto for the fruitful discussions. GMP acknowledges support by MUR under PRIN Project No. 2022FEXLYB, Quantum Reservoir Computing (QuReCo). AR acknowledges the grant ``Bando Viaggi e Soggiorni di Studio degli Studenti - Anno 2022'' from Università degli Studi di Palermo. JRP acknowledges financial support from the Spanish MICIN FPU22/01511. AS and JRP acknowledge financial support from Ministerio de Ciencia e Innovación of the Spanish Government with funding from European Union NextGenerationEU (PRTR-C17.I1) and by Generalitat de Catalunya and from  the Spanish MICIN (project PID2022-141283NB-I00) with the support of FEDER funds, and by the Ministry for Digital Transformation and of Civil Service of the Spanish Government through the QUANTUM ENIA project call - Quantum Spain project, and by the European Union through the Recovery, Transformation and Resilience Plan - NextGeneration EU within the framework of the Digital Spain 2026 Agenda.

\section*{References}
\bibliography{references}

\appendix

\section{Basic concepts of networks and percolation}
\label{sec:comp_net}

Network theory provides a single theoretical and mathematical framework to describe a broad set of complex systems, ranging from biological networks of neurons in our brain to the network of social interaction or infrastructure networks like the power grid or the internet \cite{2002albert, 2006boccaletti, 2020mini}.

A network is described by a graph $G = (V, E)$, a mathematical structure consisting of two sets: $V$, the set of vertices in the graph, called ``nodes'', and $E$, the set of ``edges'' connecting pair of nodes \cite{newman, latora}.

\begin{figure}
    \centering
    \includegraphics[width=0.7\textwidth]{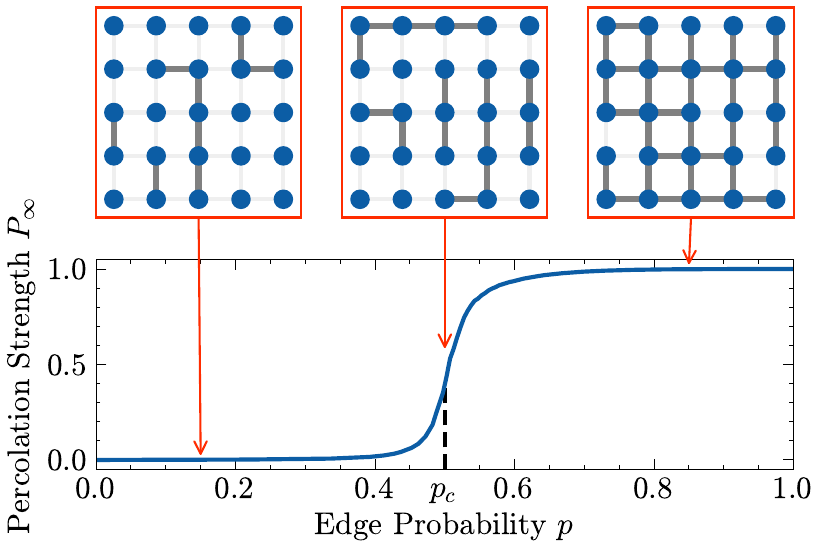}
    \caption{The plot shows an edge percolation simulation performed on a $100 \times 100$ square lattice. Three snapshots of an example of a simulation on a smaller lattice are shown for probabilities $0.15, 0.5, 0.85$. The percolation transition is highlighted with $p^\text{square}_c = 1/2$ \cite{2017hugo, 1980kesten}.}
    \label{fig:perc}
\end{figure}

The transport properties within a network or a lattice can be characterized using the mathematical framework of percolation theory \cite{2021perc}. In ``edge percolation'', a coin is tossed for each of the edges of a network and with probability $p$ the edge is kept, while with probability $1 - p$ the edge is removed. For small $p$, the majority of edges are eliminated, whereas for $p$ close to 1, most edges are left intact. The system is said to ``percolate'' if a path of edges connecting opposite sides of the network exists \cite{grimmett}.

A critical probability, known as the percolation threshold ($p_c$), defines a second order phase transition. For $p > p_c$, the system percolates, meaning that a giant connected component emerges, whereas for $p < p_c$ the system consists only of finite disjoint clusters. This transition is sharp for infinite networks (such as infinite lattices) while for finite systems you can study finite size effects at the transition point \cite{aharonystauffer}.

Defining percolation as connecting opposite sides of a network may be suitable for lattices, but it is non-trivial for more complex networks where it is not obvious what is a ``side''. Because of that, one can use different order parameters. For example, one can utilize the probability $\Pi(p)$ that a giant spanning cluster exists. In an infinite network, $\Pi(p) = 0$ for $p < p_c$, and it sharply transitions to $\Pi(p) = 1$ for $p > p_c$.

In this article, as the order parameter, we have utilized the percolation strength $P_\infty$, which is defined as the probability that a node belongs to the largest connected component (i.e., the cluster containing the largest amount of nodes in the network). As shown in \Cref{fig:perc}, below the transition, the network consists only of small finite clusters, yielding $P_\infty(p) = 0$. Above the percolation transition ($p > p_c$), the giant component emerges, but the network is generally not fully connected and small isolated clusters still exist. After the transition $P_\infty(p)$ grows continuously from $0$ to $1$ following the law $P_\infty(p) \propto (p - p_c)^\beta$ where $\beta > 0$ is the critical exponent (for example, $\beta = 5/36$ for a 2D lattice).

While $P_\infty(p)$ characterizes the macroscopic size of the connected component, understanding transport efficiency within these structures requires analyzing the internal distances between nodes. To this end, recent advances in classical network theory have generalized these foundational concepts to path-based metrics, such as shortest-path percolation on random networks \cite{2024kim}.

\section{Entanglement manipulation}
\label{sec:quant_info}

Consider two distant laboratories $A$ (Alice) and $B$ (Bob). Both are allowed to perform local quantum operations in their respective systems, and can also communicate classical information to each other. Everything that can be decomposed only with these two actions falls under the umbrella of Local Operations and Classical Communication (LOCC) \cite{2014locc}. LOCCs are the free operations that use entanglement as a resource, they cannot increase entanglement but they can transform it in various ways \cite{nielsenchuang, 2019resource}.

In 1999 Nielsen \cite{1999nielsen} demonstrated a necessary and sufficient condition to convert deterministically a bipartite entangled state $\ket{\psi}$ into a bipartite entangled state $\ket{\phi}$ under LOCC.

The theorem states the following \cite{1999nielsen}: given a state with Schmidt decomposition $\ket{\psi} = \sum_{i=1}^d \sqrt{\lambda^\psi_i} \ket{i}_A \otimes \ket{i}_B$ where the Schmidt coefficients are placed in decreasing order, i.e. $\lambda^\psi_1 \geq ... \geq \lambda^\psi_d$. It is possible to transform $\ket{\psi}$ into another bipartite state $\ket{\phi}$ using LOCC iff $\lambda^\psi = (\lambda^\psi_1, ..., \lambda^\psi_d)$ is ``majorized'' by $\lambda^\phi$, i.e. $\lambda^\psi \prec \lambda^\phi$, where \cite{bhatia, marshall, 2001review}
\begin{equation}
    \label{eq:maj}
    \lambda^\psi \prec \lambda^\phi \quad\equiv\quad \sum_{i=1}^k \lambda^\psi_i \leq \sum_{i=1}^k \lambda^\phi_i \quad \text{for each}\; k=1,...,d
\end{equation}
Nielsen theorem tells us that entanglement transformation is possible in a deterministic way using LOCC, but also that entanglement cannot be increased in this way.

For example, an operation that is allowed is entanglement distillation \cite{2000sanpera}, which starts with $N$ partially entangled states $\ket{\psi}^{\otimes N}$. These are distilled into a single, but more entangled state $\ket{\phi}$ \footnote{to be precise they are distilled into $\ket{\phi} \bigotimes_{i=1}^{N-1} \ket{\eta_i}$ where $\ket{\eta_i}$ are product states we can ignore} using LOCC as long as $\lambda^{\psi^{\otimes N}} \prec \lambda^\phi$.

Now, let's say we want to convert a single bipartite state $\ket{\psi}$ to a more entangled state $\ket{\phi}$, in this case $\lambda^\psi \nprec \lambda^\phi$ and we cannot achieve this deterministically using LOCC \cite{2014locc}.

However, this can be achieved probabilistically using Stochastic LOCC (SLOCC) (also called ``gambling with the entanglement''). In 1999 Vidal \cite{1999vidal} generalized Nielsen theorem deriving the maximum probability to convert $\ket{\psi} \to \ket{\phi}$ with the optimal SLOCC protocol
\begin{equation}
    \label{eq:vidal}
    p(\psi \to \phi) = \min_{k=1, .., d} \frac{\sum_{i=k}^d \lambda_i^\psi}{\sum_{i=k}^d \lambda_i^\phi}
\end{equation}
When $\ket{\phi} = \ket{\Psi^-}$ is a maximally entangled state of two qubits \cite{1997lopopescu}, the Schmidt coefficients are $\lambda^{\Psi^-}_1 = \lambda^{\Psi^-}_2 = 1/2$ and the Singlet Conversion Probability (SCP), that is, the probability to convert a generic 2-qubit state $\ket{\psi} \to \ket{\Psi^-}$
\begin{equation}
    \label{eq:scp}
    p(\psi \to \Psi^-) = \min\left\{\frac{\lambda^\psi_1 + \lambda^\psi_2}{\frac{1}{2} + \frac{1}{2}}, \frac{\lambda^\psi_2}{\frac{1}{2}}\right\} = \min\{1, 2\lambda_2^\psi\}
\end{equation}
The conversion of $\ket{\psi}$ into a maximally entangled Bell state (equivalently, into $\ket{\Psi^-}$ up to local transformations) can be implemented using the so-called ``Procrustean method'' \cite{1996bennet2}. Starting with the state $\ket{\psi} = \sqrt{\lambda_1} \ket{00} + \sqrt{\lambda_2} \ket{11}$ the method can be implemented with the following generalized measurement \cite{2016siomau}
\begin{equation}
    \label{eq:procrustean}
    M_1 = \begin{pmatrix}
        \sqrt{\dfrac{\lambda_2}{\lambda_1}} & 0 \\
        0 & 1
    \end{pmatrix} \qquad \qquad
    M_2 = \begin{pmatrix}
        \sqrt{1 - \dfrac{\lambda_2}{\lambda_1}} & 0 \\
        0 & 0
    \end{pmatrix}
\end{equation}
Which satisfy $M^\dagger_1 M_1 + M^\dagger_2 M_2 = \id$. The POVM outcomes are
\begin{equation}
    \label{eq:procrustean_results}
    \begin{aligned}
        \ket{\psi_1} &= \frac{(M_1 \otimes \id) \ket{\psi}}{\sqrt{\bra{\psi}(M^\dagger_1 \otimes \id)(M_1 \otimes \id) \ket{\psi}}} = \ket{\Phi^+} \\
        \ket{\psi_2} &= \frac{(M_2 \otimes \id) \ket{\psi}}{\sqrt{\bra{\psi}(M^\dagger_2 \otimes \id)(M_2 \otimes \id) \ket{\psi}}} = \ket{00}
    \end{aligned}
\end{equation}
With probabilities
\begin{equation}
    \label{eq:procrustean_probs}
    \begin{aligned}
        p(\ket{\Phi^+}) &= 2\lambda_2 \\
        p(\ket{00}) &= 1 - 2\lambda_2
    \end{aligned}
\end{equation}
Here, the state $\ket{\Phi^+} = \frac{1}{\sqrt{2}} (\ket{00} + \ket{11})$ is locally equivalent to $\ket{\Psi^-}$. The SCP $p_\psi$ is then defined as $p(\ket{\Phi^+}) = p(\ket{\Psi^-})$. With probability $p(\ket{00}) = 1 - p_\psi$, instead, we are left with no entanglement at all.

This SCP can also be understood as an entanglement measure: maximally entangled states have an SCP of $1$ while separable states have an SCP of $0$.

\section{Multigraph quantum networks}
\label{sec:multi}

The initial setup to implement the $q$-swap operation described in \Cref{sec:qep} is a network sharing multiple bipartite entangled states between users, i.e. a ``Multigraph quantum network'' \cite{newman}.

\begin{figure}
    \centering
    \includegraphics[width=0.5\textwidth]{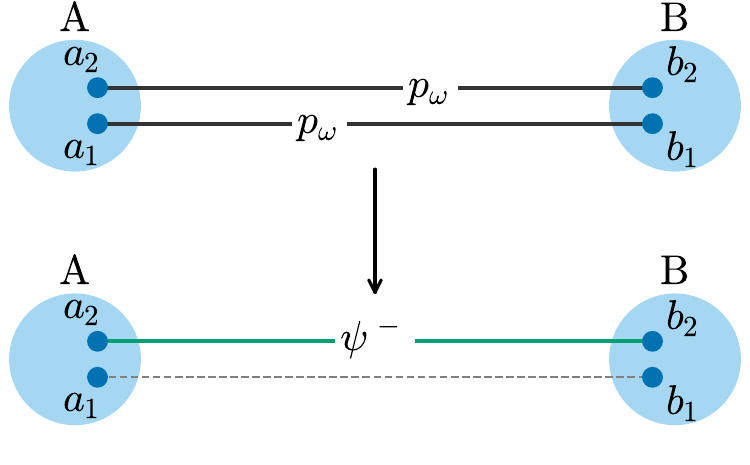}
    \caption{An example of a successful SLOCC protocol for a multiedge. Here, two identical states $\ket{\omega}$ are converted into a single maximally entangled state, the remaining state is just a product state.}
    \label{fig:multi-slocc}
\end{figure}

Starting with two identical states $\ket{\omega}$ one can apply the SLOCC purification method like in \Cref{fig:multi-slocc} in multiple ways:
\begin{itemize}
    \item One can convert the two initial states separately and ask what is the probability that at least one is successful. Then we will have that $p_\text{sep} = 1 - (1 - p_\omega)^2 = 2p_\omega - p_\omega^2$ \cite{2008perseguers}.
    \item We can implement the optimal SCP from \Cref{eq:vidal}, with probability
    \begin{equation}
        \label{eq:double}
        p(\omega \otimes \omega \to \Psi^-) = 2(1 - (\lambda^\omega_1)^2) = 2(1 - (1 - \lambda^\omega_2)^2) = 2p_\omega - \frac{p_\omega^2}{2}
    \end{equation}
    We will call this 2-state distillation SCP $p_{\omega^{\otimes 2}}$. This result is generalized to the case of $N$-state distillation as $p_{\omega^{\otimes N}} = 2(1 - (\lambda^\omega_1)^N)$.
\end{itemize}
For example, if we want to calculate the CEP for a double-bond network then we will have that $p_{\omega^{\otimes 2}} = p_c$. So, we get that the minimum amount of entanglement for each copy of the initial state is
\begin{equation}
    \label{eq:double-cep}
    p^\text{CEP}_\text{2-network} = 2-\sqrt{4 - 2 p_c}
\end{equation}
This, as expected, is always lower than the CEP for a single network which is just $p^\text{CEP}_\text{1-network} = p_c$ (less initial entanglement per state is needed).

\section{Multigraph random quantum networks}
\label{sec:rand-multi}

As in the case of entanglement percolation in standard networks, we are still allowed to have multiple states from one node to another as in \Cref{eq:double}. We will now have a ``Multigraph Random quantum network''. With this new model we have to distinguish two possible cases:

\begin{figure}
    \centering
    \begin{subfigure}[b]{0.39\textwidth}
        \includegraphics[width=\textwidth]{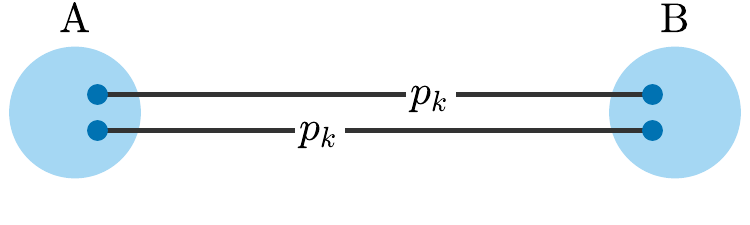}
        \caption{Equal random multiedge.}
        \label{fig:repeated-multi}
    \end{subfigure}
    \hspace{4em}
    \begin{subfigure}[b]{0.39\textwidth}
        \includegraphics[width=\textwidth]{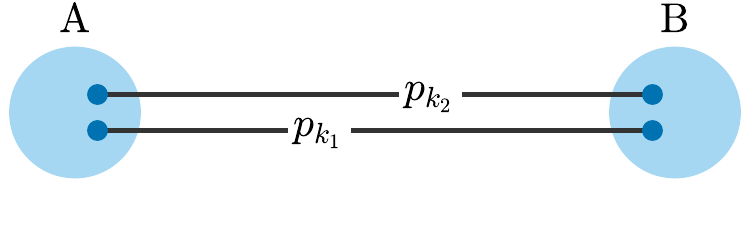}
        \caption{Independent random multiedge.}
        \label{fig:random-multi}
    \end{subfigure}
    \caption{Examples of a multiedge in a random state network. Here, two edges connect node $A$ with node $B$. (a) For ``equal multiedge'' we will have that only one probability $p_k$ is drawn from the distribution and assigned to all the edges inside a multiedge. (b) For ``independent multiedge'' we have that all the edges inside a multiedge are drawn from the distribution.}
    \label{fig:multiedge}
\end{figure}

\begin{itemize}
    \item \textbf{Equal multiedge:} all the (random) edges connecting the same two users are equal, as shown in \Cref{fig:repeated-multi}. Here, the SLOCC purification probability will be
    \begin{equation}
        p(\psi_k^{\otimes 2} \to \Psi^-) = p_{k^{\otimes 2}} = \min\left\{1, 2\left[1 - \left(\lambda_1^{\psi_k}\right)^2\right]\right\}
        \label{eq:repeated-multi}
    \end{equation}
    \item \textbf{Independent multiedge:} all the edges are independently drawn from the distribution of the SCPs, as shown in \Cref{fig:random-multi}. The SLOCC probability in this case is
    \begin{equation}
        p(\psi_{k_1} \otimes \psi_{k_2} \to \Psi^-) = p_{k_1 \otimes k_2} = \min\left\{1, 2\left(1 - \lambda_1^{\psi_{k_1}} \lambda_1^{\psi_{k_2}}\right)\right\}
        \label{eq:random-multi}
    \end{equation}
\end{itemize}
We can write both these equations as a function of the SCP $p_k = 2\lambda_2^{\psi_k}$ in the following way
\begin{equation}
    \begin{aligned}
        p_{k^{\otimes 2}} &= 2p_k - \frac{p_k^2}{2} \\
        p_{k_1 \otimes k_2} &= p_{k_1} + p_{k_2} - \frac{p_{k_1} p_{k_2}}{2}
    \end{aligned}
\end{equation}
We can then average these results out to obtain
\begin{equation}
    \begin{aligned}
        \langle p_{k^{\otimes 2}} \rangle &= 2\langle p \rangle - \frac{\langle p^2 \rangle}{2} \\
        \langle p_{k_1 \otimes k_2} \rangle &= 2\langle p \rangle - \frac{\langle p \rangle^2}{2}
    \end{aligned}
    \label{eq:averages}
\end{equation}
We will always have $\langle p^2 \rangle \geq \langle p \rangle^2$ so at the end we conclude that
\begin{equation}
    \langle p_{k_1 \otimes k_2} \rangle \geq \langle p_{k^{\otimes 2}} \rangle
\end{equation}
The average SLOCC probability of having the multiple states all randomly distributed is always higher than the case of repeated multiedges. Therefore, we conclude that the random multiedge is always better than the repeated multiedge.

\section{RCEP is the CEP of the average}
\label{sec:rcep-theory}

The results in \Cref{sec:rand-cep} were also obtained for different initial network topologies (Erdős–Rényi, Barabási-Albert and Watts-Strogatz networks) and for different distributions of the SCPs (truncated Gaussian and bimodal distributions). Moreover, this result is consistent with recent findings in dynamical entanglement percolation where in the absence of spatial correlations, the macroscopic order parameter collapses onto that of uniform bond percolation \cite{2026cirigliano}.

Theoretically, one can prove \Cref{eq:random-cep} in the following way:
\begin{itemize}
    \item For standard percolation one starts with a network of $n$ edges, with probability $p$ for each edge. The edges are either empty or occupied, with $\omega_i = 0$ indicating a removed edge, while $\omega_i = 1$ indicates an edge that is kept intact.

    A percolation trial leaves the network with $r$ remaining edges where $r = \sum_i \omega_i$, constituting a specific microstate $\omega$. The probability of getting this specific microstate is the same as independent Bernoulli trials:
    \begin{equation}
        P(\omega) = p^{r}(1-p)^{n - r}
    \end{equation}

    \item For random percolation each edge $i$ now has a probability $p_i$, in general, drawn from a distribution $f(p)$. Given the probabilities, the probability to get a specific microstate is
    \begin{equation}
        P(\omega|\{p_i\}) = \prod_{i=1}^n p_i^{\omega_i} (1-p_i)^{1-\omega_i}
    \end{equation}
    We can then average out over the distribution of probabilities to get
    \begin{equation}
        \begin{aligned}
            \overline{P(\omega)} &= \int_0^1\dots\int_0^1 \left[ \prod_{i=1}^n p_i^{\omega_i} (1-p_i)^{1-\omega_i} \right] \prod_{i=1}^n f(p_i) dp_1 \dots dp_n = \\
            &= \prod_{i=1}^n \int_0^1 p_i^{\omega_i} (1-p_i)^{1-\omega_i} f(p_i) dp_i = \\
            &= \prod_{i=1}^n I(\omega_i)
        \end{aligned}
    \end{equation}
    We can then calculate the integral for the case when the edge is kept or removed
    \begin{equation}
        \begin{aligned}
            I(0) &= \int_0^1 (1 - p_i) f(p_i) dp_i = 1 - \langle p \rangle \\
            I(1) &= \int_0^1 p_i f(p_i) dp_i = \langle p \rangle
        \end{aligned}
    \end{equation}
    Which means that, in general, the integral is
    \begin{equation}
        I(\omega_i) = \langle p \rangle^{\omega_i} (1 - \langle p \rangle)^{1 - \omega_i}
    \end{equation}
    By using the fact that $r = \sum_i \omega_i$ we then can get that
    \begin{equation}
        \overline{P(\omega)} = \langle p \rangle^r (1 - \langle p \rangle)^{n - r}
    \end{equation}
\end{itemize}
This means that, because both models (RCEP and CEP of the average) generate the same configurations with the same probabilities, their macroscopic topological features must be identical, hence, percolation properties are the same.

\end{document}